\documentclass[a4paper, 10pt, conference]{ieeeconf}      

\IEEEoverridecommandlockouts                              

\usepackage{amsfonts}
\usepackage{amsmath}         
\usepackage{hyperref}
\usepackage{cite}
\usepackage{cleveref}
\usepackage{color}
\usepackage{algorithm}
\usepackage{algorithmic}
\let\labelindent\relax
\usepackage{enumitem}
\usepackage{float}
\usepackage{graphicx}
\usepackage{booktabs}
\usepackage{threeparttable}
\usepackage{subfigure}

\newcommand{\diag}{\mathop{\rm diag}\nolimits}

\begin{document}
\title{\LARGE \bf
Equivalence of SS-based MPC and ARX-based MPC}
\author{Liang Wu$^{1}$%
\thanks{The authors are with the IMT School for Advanced Studies Lucca, Italy,
        {\tt\small liang.wu\@imtlucca.it}}%
}
\thispagestyle{empty}
\pagestyle{empty}
\maketitle

\begin{abstract}
Two kinds of control-oriented models used in MPC are the state-space (SS) model and the input-output model (such as the ARX model). The SS model has interpretability when obtained from the modeling paradigm, and the ARX model is black-box but adaptable. This paper aims to introduce interpretability into ARX models, thereby proposing a first-principle-based modeling paradigm for acquiring control-oriented ARX models, as an alternative to the existing data-driven ARX identification paradigm. That is, first to obtain interpretative SS models via linearizing the first-principle-based models at interesting points and then to transform interpretative SS models into their equivalent ARX models via the SS-to-ARX transformations. This paper presents the Cayley-Hamilton, Observer-Theory, and Kalman Filter based SS-to-ARX transformations, further showing that choosing the ARX model order should depend on the process noise to achieve a good closed-loop performance rather than the fitting criteria in data-driven ARX identification paradigm. An AFTI-16 MPC example is used to illustrate the equivalence of SS-based MPC and ARX-based MPC problems and to investigate the robustness of different SS-to-ARX transformations to noise.

\end{abstract}


\section{Introduction}\label{sec:intro}
Model Predictive Control (MPC) is a successful feedback control method to control multi-input multi-output (MIMO) systems subject to constraints \cite{morari1999model}. MPC has been widely used in diverse industrial areas, such as process \cite{qin2003survey,bemporad1999robust}, energy management \cite{chen2013mpc,di2013stochastic}, aerospace \cite{eren2017model,guiggiani2015fixed}, and power electronics\cite {geyer2016model,cimini2015online}. The critical component for a successful MPC controller is obtaining a simplified control-oriented prediction model of the physical process to predict its likely evolution. Two main kinds of control-oriented models used in MPC are the state-space (SS) and input-output model. Most earlier MPC methods employ a different type of input-output models, like the Model Predictive Heuristic Control (MPHC) \cite{richalet1978model}, the Model Algorithmic Control (MAC) \cite{ROUHANI1982401}, the Internal Model Control (IMC) \cite{garcia1982internal}  the Dynamic Matrix Control (DMC) \cite{cutler1980dynamic}, and the Generalized Predictive Control (GPC) \cite{CLARKE1987137, CLARKE1987149}. Input-output models were almost superseded by state-space models after three decades of MPC development \cite{mayne2014model}, and the SS model provides a series of mature theories to address controllability, observability, optimal control, etc., elegantly \cite{kalman1960contributions,kalman1970lectures}. Furthermore, a SS model can be regarded as interpretative when obtained from the first-principle-based modeling paradigm. In contrast, the input-output model, such as the Auto-Regressive model with eXogenous (ARX) terms model, is the black-box model without interpretability \cite{ljung2001black}. 

Nonetheless, input-output models are still prevalent and preferred in industrial MPC applications. One reason is that most industrial control application scenarios cannot measure full state information, which is required by state-space-based (SS-based) MPC. In that context, an observer (or estimator), like the well-known Luenberger observer, the Kalman Filter, and the moving horizon estimator \cite{rawlings2006particle}, is adopted to estimate full states from streaming input-output data. The simultaneous deployment of an observer and MPC not only increases tuning difficulties and online computation costs when the dimension of states is large. Clearly, an input-output ARX model-based (ARX-based) MPC does not require cooperating with an online observer. Another reason is that the input-output ARX model is widely used in adaptive control \cite{aastrom2013adaptive}, and its cheap online adaptation cost is attributed to the linear relationship between input and output. The adaptability of input-output ARX models is very appealing in practical industrial scenarios where practical plants suffer changing dynamics in their whole life operation cycle, such as in the case of fouling of heating equipment in chemical processes, changes of mass and inertia in rockets due to fuel consumption, and many others.

The input-output ARX model has long been considered black-box because its acquisition is through a data-driven paradigm such as system identification methods. Conversely, the SS model can be obtained either through the first-principle-based modeling paradigm or through the data-driven paradigm. The first-principle-based modeling paradigm typically builds nonlinear SS models with interpretability, which can derive simplified control-oriented models for MPC via linearization on operating ranges or online successive linearization \cite{wu2022rapid}. Regarding adaptability, the SS model is inferior compared to the ARX model. There are two approaches for the SS model to achieve adaptability: one is that when the SS model comes from the first-principle-based modeling paradigm, the joint estimation of model parameters and states can be adopted via the dual Extended Kalman Filter (EKF) \cite{wenzel2006dual}; the other is that when the SS model comes from the data-driven paradigm, the online subspace identification algorithm, like Multivariable Output Error State Space (MOESP) \cite{verhaegen1994identification} and Numerical algorithms for Subspace State Space System Identification (N4SID) algorithms \cite{van1994n4sid}, can be adopted. Both are more computationally heavy and complicated than the online ARX identification method. The subspace methods of the latter are based on robust numerical tools such as QR decomposition and singular value decomposition, which are not suitable for online schemes. An online updating scheme of the MIMO ARX model can be decomposed into some multi-input-single-output models, which then can be updated parallelly by recursive-least-squares (RLS) or Kalman Filter (KF) algorithms without matrix-inverse operations. It seems difficult to preserve interpretability and adaptability simultaneously in a control-oriented model from a computational perspective.

The relationship between SS models and ARX models has been investigated. An ARX model can be transformed into an SS model, and vice versa. Most literature treats an ARX-based MPC problem as a standard SS-based MPC problem by converting an ARX model into an SS model, see \cite{huusom2010arx}. In \cite{aoki1991state}, Aoki and Havenner only revealed that an  SS model can be transformed into an ARX model by using the Cayley-Hamilton theorem \cite{gantmakher1959theory}, without further discussing its application in MPC. In \cite{phan1970relationship,phan1993linear}, Minh et al. found that an SS model can be transformed into an equivalent ARX model in terms of an observer gain, thus proposing their data-driven predictive control scheme in which the control input was a linear combination of past input-output data \cite{phan1996predictive, lim1997identification, phan1998unifying}, but without introducing constraints capabilities like modern MPC technology. Few works explored other application values of the SS-to-ARX transformation. 

This paper leverages equivalent SS-to-ARX transformations to introduce interpretability into an adaptive ARX model.
The aim is to propose the first-principle-based modeling paradigm for acquiring control-oriented ARX models, as an alternative to the existing data-driven ARX identification paradigm. Since numerous rigorous mechanistic (first-principle-based) model of the plant already exists, it would be attractive to take advantage of those models in an MPC algorithm. However, their drawback is a huge computational burden even if using their linearized SS models in a real-time MPC setting, when they involve a large state dimension, such as in distributed parameter systems.  Clearly, an input-output ARX model is compact and can tackle the issue of large state dimensions. But the data-driven ARX identification paradigm relies on simulation data from a rigorous mechanistic model to obtain an ARX model with the loss of its interpretability, and our proposed modeling ARX paradigm analytically derives the ARX model thus inheriting the interpretability of the rigorous mechanistic model. Simultaneously persevering the interpretability and adaptability is the desired solution for safety-critical and adaptive MPC applications. 

The proposed first-principle-based modeling ARX acquiring paradigm first obtains an (or multiple) interpretative SS model at the operating range via linearizing the existing first-principle-based model and then transforms the interpretative SS model into its equivalent ARX model via the SS-to-ARX transformations. This paper presents the Cayley-Hamilton (CH), Observer-Theory (OT), and Kalman Filter (KF) based SS-to-ARX transformations to show that choosing the ARX model order should depend on the process noise to achieve a good closed-loop performance, rather than the fitting
criteria in data-driven ARX identification paradigm. Note that the interpretative ARX model can adopt an online updating scheme to handle time-varying dynamics in the ARX-based MPC problems, which is an interpretive and adaptive MPC framework in our related work \cite{wu2022interpretative}.

\subsection{Structure and Notation}
The structure of the paper is as follows. Section \ref{sec:probelm_description} first defines the SS-based MPC tracking problem in its subsection \ref{sec:SS-based MPC}, and subsections \ref{sec:CH-based SS-to-ARX transformation}, \ref{sec:observer-theory-based transformation}, and \ref{sec:KF-based SS-to-ARX transformation} present the derivations of the CH-based, OT-based, and KF-based SS-to-ARX transformations, respectively. Section \ref{sec:ARX-based MPC} shows the corresponding ARX-based MPC tracking problem and mentions its solving algorithm. In Section \ref{sec:example}, numerical experiments are presented. Finally, we draw conclusions in Section \ref{sec:conclusion}.

$H \succ 0$ ($H \succeq 0$) denotes positive definiteness (semi-definiteness) of a square matrix $H$. For a vector $z$, $\|z\|_{H}^{2}$ denotes the operation $z^{\prime}Hz$. $H^{\prime}$ (or $z^{\prime}$) denotes the transpose of matrix $H$ (or vector $z$).

\section{Problem description and methods}\label{sec:probelm_description}
\subsection{State-space based MPC problem}\label{sec:SS-based MPC}
This paper considers the following discrete-time state-space model,
\begin{subequations}\label{SS}
\begin{eqnarray}
x_{t+1}&=& A x_{t}+B u_{t}\label{SS_x}\\
y_{t} &=& C x_{t}\label{SS_y}
\end{eqnarray}
\end{subequations}
where each $\{x_t\} \in \mathbb{R}^{n}$ are the state variables, each $\{u_t\} \in \mathbb{R}^{q}$ are the input variables and each $\{y_t\} \in \mathbb{R}^{m}$ are the output variables. Then, an SS-based MPC tracking problem formulation is shown as follows,
\begin{eqnarray}
\min && \sum_{t=0}^{T-1}\left\| \left(y_{t+1}-r_{t+1}\right)\right\|_{W_y}^{2} + \left\|\Delta u_{t}\right\|_{W_{\Delta u}}^{2} \nonumber\\
\text {s.t.}  && x_{t+1}=A x_{t}+B u_{t},t=0,\ldots,T-1\nonumber\\
&& y_{t} = C x_{t}, t=1,\ldots,T \nonumber\\
&& u_{t} = u_{t-1} + \Delta u_{t}, t=0,\ldots,T-1\nonumber\\
&& y_{\min } \leq y_{t} \leq y_{\max },\ t=1, \ldots, T\nonumber\\
&& u_{\min } \leq u_{t} \leq u_{\max },\ t=0, \ldots, T-1\nonumber\\
&& \Delta u_{\min } \leq \Delta u_{t} \leq \Delta u_{\max }, t= 0, \ldots, T-1\nonumber\\
&& x_{0} = \hat{x}_0
\label{SS_MPC}
\end{eqnarray} 
where $\{r_{t}\}$ are the desired output tracking reference signals, $\{y_{t}\}$ are the measured outputs subject to bound constraints $[y_{\min},y_{\max}]$, $\{u_{t}\}$ are the control inputs subject to bound constraints $[u_{\min},u_{\max}]$, $\{\Delta u_{t}\}$ are control input increments subject to bound constraints $[\Delta u_{\min},\Delta u_{\max}]$, $W_y\succ 0$ and $W_{\Delta u}\succ 0$ are the diagonal weights matrices. $\hat{x}_0$ is the estimated value of full state variable at current time $t$, which is estimated by estimation algorithm or observer. We assume that only the measured outputs, inputs, and input increments have box constraints; the states do not.

\subsection{Cayley-Hamilton based SS-to-ARX transformation}\label{sec:CH-based SS-to-ARX transformation}
Based on the states evolution equation (\ref{SS_x}), calculating from previous time $t-n$ time to current time $t$ can lead to the following equation
\begin{equation}
\begin{aligned}
x_{t} & = A^{n} x_{t-n} \\
& + {\left[\begin{array}{r}
B,AB,\cdots,A^{n-1}B
\end{array}\right]}
{\left[\begin{array}{c}
u_{t-n} \\
u_{t-n+1} \\
\vdots \\
u_{t-1}
\end{array}\right]}
\end{aligned}
\end{equation}
Writing down each output equation  (\ref{SS_y}) from previous time $t-n$ to current time $t$,
\begin{equation}\label{y_evolution}
\begin{aligned}
y_{t-n} &= C x_{t-n} \\
y_{t-n+1} &= C Ax_{t-n} + CBu_{t-n}  \\
y_{t-n+2} &= C A^2x_{t-n} + CBu_{t-n+1} + CABu_{t-n} \\
\vdots \\
y_{t} & = CA^nx_{t-n}\\
&+{\left[\begin{array}{r}
CB,CAB,\cdots,CA^{n-1}B
\end{array}\right]}
{\left[\begin{array}{c}
u_{t-n} \\
u_{t-n+1} \\
\vdots \\
u_{t-1}
\end{array}\right]}
\end{aligned}
\end{equation}
The Cayley-Hamilton theorem \cite{gantmakher1959theory} is introduced to remove the state vector $x_{t-n}$. The Cayley–Hamilton theorem states that every square matrix satisfies its own characteristic equation. For a given $n \times n$ matrix $A$, then the characteristic polynomial of $A$ is defined as $p_A(\lambda) = \operatorname{det}(\lambda I_n - A)$, the determinant is also a degree-n monic polynomial in $\lambda$, $p_{A}(\lambda)=\lambda^{n}+c_{1} \lambda^{n-1}+\cdots+c_{n-1} \lambda+c_{n}$. One can create an analogous polynomial $p_{A}(A)$ in the matrix $A$ instead of the scalar variable $\lambda$, defined as $p_{A}(A)=A^{n}+c_{1} A^{n-1}+\cdots+c_{n-1} A+c_{n} I_{n}$. The Cayley–Hamilton theorem states that this polynomial expression is equal to the zero matrices, which allows $A^n$ to be expressed as a linear combination of the lower matrix powers of A.
\begin{equation}\label{CH_theorem}
A^{n}+c_{1} A^{n-1}+\cdots+c_{n-1} A+c_{n} I_{n}=\left(\begin{array}{ccc}
0 & \cdots & 0 \\
\vdots & \ddots & \vdots \\
0 & \cdots & 0
\end{array}\right)
\end{equation}
By substituting this equation (\ref{CH_theorem}) into the output equations (\ref{y_evolution}) from previous $t-n$ time to current time $t$, we can obtain the following transformed ARX model
\begin{equation}
\begin{aligned}
&y_{t}+c_{1}y_{t-1}+\cdots+c_{n-1}y_{t-n+1}+c_{n}y_{t-n}\\
&= \Theta_{1} u_{t-1}+ \Theta_{2} u_{t-2} + \Theta_{3} u_{t-3}+\ldots+\Theta_{n} u_{t-n}
\end{aligned}
\label{ARX}
\end{equation} 
where $\Theta_{1} = CB, \Theta_{2}=CAB+c_{1}CB, \Theta_{3}=CA^2B+c_{1}CAB+c_{2}CB, \cdots, \Theta_{n} = CA^{n-1}B+c_{1}CA^{n-2}B+\cdots+c_{n}CB$.

The CH-based SS-to-ARX transformation implies that the order of the ARX model is equal to the state dimension of the SS model. However, the closed-loop performance of its corresponding ARX-based MPC is sensitive to process noise and measurement noise. Because the corresponding ARX-based MPC is essentially equivalent to the SS-based MPC controller with a deadbeat observer in that case. A deadbeat observer is a minimum-time observer which is sensitive to process noise and measurement noise. The next subsection \ref{sec:observer-theory-based transformation} shows that the observed-theory-based (OT-based) SS-to-ARX transformation is a generalized transformation theory, which can be robust to noise.

\subsection{Observer-Theory based SS-to-ARX transformation}\label{sec:observer-theory-based transformation}
Begin deriving the SS-to-ARX transformation utilizing a gain matrix $L$ in the SS model~(\ref{SS}) as follows
\begin{equation}\label{SS_xyu}
\begin{aligned}
    x_{t+1} &= A x_t + B u_t  - L y_t + L y_t\\
    &=(A-LC)x_t+B u_t + L y_t
\end{aligned}
\end{equation}
Based on the evolution equation (\ref{SS_xyu}), calculating from previous time $t-p$ to current time $t$ can lead to the following equation
\begin{equation}\label{SS_p_t}
\begin{aligned}
    x_{t}&=(A-LC)^p x_{t-p} +\sum_{i=1}^{p}(A-LC)^{i-1} L y_{t-i} \\ 
    &+ \sum_{i=1}^{p}(A-LC)^{i-1} B u_{t-i}
\end{aligned}
\end{equation}
If a gain matrix $L$ exists such that $(A-LC)^p$ vanishes to zero, i.e.,
\begin{equation}\label{condition_ALC}
(A - LC)^{k} \equiv 0, \quad k \geq p
\end{equation}
From linear system theory, the existence of such a gain matrix $L$ is assured as long as the system is observable. Thus, the term $(A-LC)^p x_{t-p}$ in (\ref{SS_p_t}) is zero for $k\geq p$, then multiplying the matrix $C$ on both sides of the equation (\ref{SS_p_t}) can derive the following ARX model,
\begin{equation}\label{OT-based SS-to-ARX}
\begin{aligned}
    y_{t} &= \sum_{i=1}^{k}C(A-LC)^{i-1} L y_{t-i}\\
    &+ \sum_{i=1}^{k}C(A-LC)^{i-1} B u_{t-i}
\end{aligned}
\end{equation}

Next, we provide the analysis to show that the gain matrix $L$ can be viewed as an observer gain. The original state-space model~(\ref{SS}) has an observer gain $L$ of the following form
\begin{equation}\label{OT_SS_to_ARX}
\begin{aligned}
\hat{x}_{t+1} &= A\hat{x}_t + Bu_t + L(y_t-\hat{y}_t)\\
\hat{y}_t &= C\hat{x}_t
\end{aligned}
\end{equation}
where $\hat{x}_t$ is the estimated state. The state estimation error can be denoted as $e_t=x_t-\hat{x}_t$, then $e_t$ follows the dynamic equation,
\begin{equation}
    e_{t+1} = (A-LC)e_t
\end{equation}
Thus, the estimated state $\hat{x}_t$ will converge the actual value $x_t$ as $t$ tends to infinity if the matrix $A-LC$ is asymptotically stable, namely the condition (\ref{condition_ALC}). In fact, the connection between CH-based and OT-based SS-to-ARX transformation can be established by a generalized Cayley-Hamilton theorem.
Defining a sequence matrices $\{M_i\}$, each of them belongs to $\mathbb{R}^{n\times m}$. As long as $k m\geq n$ then it is guaranteed for an observable system that $\{M_i\}$ exists such that
\begin{equation}
    A^p+M_1 CA^{p-1} + M_2 CA^{p-2} + \ldots + M_p C = 0
\end{equation}
which is a generalized Cayley-Hamilton theorem. Following the same idea of the standard Cayley-Hamilton-based derivation in section \ref{sec:CH-based SS-to-ARX transformation}, can derive a general ARX model as follows
\begin{equation}
y_{t}=\sum_{i=1}^{p}\left(\bar{A}_{i} y_{t-i}+\bar{B}_{i} u_{t-i}\right)
\end{equation}
We omit this detailed derivation since the OT-based SS-to-ARX transformation is more easily computable, only required to satisfy the condition (\ref{condition_ALC}) such as using pole placement methods.

\subsection{Kalman Filter based SS-to-ARX transformation}\label{sec:KF-based SS-to-ARX transformation}
This subsection shows the derivation of the SS-to-ARX transformation in the presence of process and measurement noise. It tells that an ARX model can be equivalent to an SS model with its optimal Kalman filter. Consider the case where the state-space model~(\ref{SS}) has process and measurement noise
\begin{equation}\label{Stochastic-SS}
\begin{aligned}
x_{t+1}&=A x_{t}+B u_{t}+w_t\\
y_{t} &=C x_{t}+v_t
\end{aligned}
\end{equation}
where the process noise $w_t$ and measurement noise $v_t$ are two zero mean white noise with covariances Q and R, respectively. The linear stochastic state-space~(\ref{Stochastic-SS}) model can also be expressed in the form of a Kalman filter
\begin{equation}\label{Stochastic-SS-Kalman}
\begin{aligned}
\hat{x}_{t+1}&=A \hat{x}_{t}+B u_{t}+K\epsilon_t\\
y_{t} &=C \hat{x}_{t}+\epsilon_t
\end{aligned}
\end{equation}
where $\epsilon_t$ is a white sequence of residual with covariance $\Sigma = CPC^{\prime}+R$ and $P$ are the unique positive definite symmetric solution of the algebraic Riccati equation
\begin{equation*}
P=APA^{\prime}-APC^{\prime}(CPC^{\prime}+R)^{-1}CPA^{\prime}+Q
\end{equation*}
The Kalman filter gain $K$ is given by $K=APC^{\prime}(CPC^{\prime}+R)^{-1}=APC^{\prime}(\Sigma)^{-1}$. Then, the system~(\ref{Stochastic-SS-Kalman}) can be expressed as
\begin{equation}\label{Stochastic-SS-Kalman-y}
\begin{aligned}
\hat{x}_{t+1}&=A \hat{x}_{t}+B u_{t}+K\epsilon_t-Ky_t+Ky_t\\
&=(A-KC)\hat{x}_{t}+B u_{t}+K \epsilon_t-K \epsilon_t+Ky_t\\
&=(A-KC)\hat{x}_{t}+B u_{t}+Ky_t\\
y_t &= C\hat{x}_t+\epsilon_t
\end{aligned}
\end{equation}
Based on the above evolution equation (\ref{Stochastic-SS-Kalman-y}), calculating from previous time $t-p_1$ to current time $t$ can lead to the following equation
\begin{equation}\label{p1-input-output}
\begin{aligned}
    y_{t}&=C(A-KC)^{p_1}\hat{x}_t+\sum_{i=1}^{p_1}C(A-KC)^{i-1}Ky_{t-i}\\
    &+\sum_{i=1}^{p_1}C(A-KC)^{i-1}Bu_{t-i}+\epsilon_{t}
\end{aligned}
\end{equation}
Provided that $p_1$ is enough large such that 
\begin{equation*}
    (A-KC)^{k}\approx0, k\geq p_1
\end{equation*}
then the input-output model~(\ref{p1-input-output}) can be approximated by the following ARX model
\begin{equation}\label{p1-input-output-approx}
\begin{aligned}
y_{t}&=\sum_{i=1}^{p_1}C(A-KC)^{i-1}Ky_{t-i}\\
&+\sum_{i=1}^{p_1}C(A-KC)^{i-1}Bu_{t-i}+\epsilon_{t}
\end{aligned}
\end{equation}
which has the same structure as equation  (\ref{OT_SS_to_ARX}) of the deterministic case, only that the value of $p_1$ may be larger than $p$. This shows that the equation (\ref{OT_SS_to_ARX}) is also robust to noise, and its noise robustness increases as the order increases. More importantly, it tells that directly online identifying an ARX model can replace the design procedure of a Kalman Filter in which the process and measurement noise covariances require to be estimated. 

\section{ARX-based MPC problem}\label{sec:ARX-based MPC}
After the above discussion, an interpretative SS model can be equivalently transformed into an adaptive ARX model. Thus, the SS-based MPC problem (\ref{SS_MPC}) has the following corresponding equivalent ARX-based MPC problem
\begin{eqnarray}
\min && \sum_{k=0}^{T-1}\left\|\left(y_{t+1}-r_{t+1}\right)\right\|_{W_y}^{2} + \left\|\Delta u_{t}\right\|_{W_{\Delta u}}^{2} \nonumber\\
\text {s.t.} &&y_{t}=\sum_{i=1}^{p}(\overline{A}_i y_{t-i} + \overline{B}_{i}u_{t-i}),t=1,\ldots,T\nonumber\\
&& u_{t} = u_{t-1} + \Delta u_{t},t=0,\ldots,T-1\nonumber\\
&& y_{\min } \leq y_{t+1} \leq y_{\max },t=0,\ldots,T-1\nonumber\\
&& u_{\min } \leq u_{t} \leq u_{\max },t=0,\ldots,T-1\nonumber\\
&& \Delta u_{\min } \leq \Delta u_{t} \leq \Delta u_{\max }, t= 0, \ldots, T-1
\label{ARX_MPC}
\end{eqnarray} 
where the past input-output data $\{u_{-1},\cdots,u_{1-p}\}$, $\{y_{0},y_{-1}\cdots,y_{1-p}\}$ needs to be provided.

Traditionally, solving the ARX-based MPC problem~(\ref{ARX_MPC}) has first to be constructed into a condensed Quadratic Programming (QP) problem or a sparse QP problem via the construction procedure, and then apply some standard QP algorithms to obtain a solution. The construction procedure can be done once offline when an ARX model would not update online. Conversely, the construction procedure must be performed online when the ARX model updates online. Thus, the computation cost of the online construction procedure and the QP solving procedure should be considered together. In some cases, their computation cost is comparable because solving the QP problem itself is very cheap thanks to the warm-starting, gradually changing set-points, or slowly changing dynamics \cite{wu2021simple}. Our previous work  \cite{wu2022construction} proposed a construction-free ARX-baed MPC algorithm suitable for our considered application scenarios. And it has presented the principle of the construction-free ARX-baed MPC algorithm and the numerical comparisons with other non-construction-free algorithms. Thus, this study applies our construction-free ARX-baed MPC algorithm to solve the ARX-based MPC problem (\ref{ARX_MPC}) and not provides numerical comparisons with other algorithms due to limited space. Moreover, the primary purpose of this study is to present the equivalence of SS-based MPC and ARX-based MPC and investigate the robustness of different SS-to-ARX transformations to noise via closed-loop simulations.

\section{Numerical example}\label{sec:example}
This section utilizes a classic open-loop unstable AFTI-16 aircraft example to show the equivalence of SS-based MPC and ARX-based MPC and investigate the sensitivity of different SS-to-ARX transformations to the process and measurement noise. The reported comparison simulation results were obtained on a MacBook Pro with 2.7~GHz 4-core Intel Core i7 and 16GB RAM, and ARX-based MPC problems are solved by our C-mex implementation of \cite{wu2022construction} in MATLAB (2020a).

The two-input-two-output plant is the continuous-time linearized AFTI-16 aircraft model reported in \cite{KAS88,bemporad1997nonlinear} as follows
\[
\left\{\begin{aligned}
\dot{x} =&{\footnotesize\left[\begin{array}{cccc}
-0.0151 & -60.5651 & 0 & -32.174 \\
-0.0001 & -1.3411 & 0.9929 & 0 \\
0.00018 & 43.2541 & -0.86939 & 0 \\
0 & 0 & 1 & 0
\end{array}\right]} x\\&+{\footnotesize\left[\begin{array}{cc}
-2.516 & -13.136 \\
-0.1689 & -0.2514 \\
-17.251 & -1.5766 \\
0 & 0
\end{array}\right] }u \\
y =&{\footnotesize\left[\begin{array}{llll}
0 & 1 & 0 & 0 \\
0 & 0 & 0 & 1
\end{array}\right]x}
\end{aligned}\right.
\]
The sampling time of designing a digital controller is $0.05$~s, and the corresponding discrete-time model using exact discretization is
\[
\left\{
\begin{aligned}
x_{t+1} &={\tiny\left[\begin{array}{cccc}
0.0093 & -3.0083 & -0.1131 & -1.6081 \\
4.7030\times 10^{-6} & 0.9862 & 0.0478 & 3.8501\times 10^{-6} \\
3.7028\times 10^{-6} & 2.0833 & 1.0089 & -4.3616\times 10^{-6} \\
1.3556\times 10^{-7} & 0.0526 & 0.0498 & 1
\end{array}\right]}x_t\\
&+{\tiny\left[\begin{array}{cc}
-0.0804 & -0.6347 \\
-0.0291 & -00143 \\
-0.8679 & -0.0917 \\
-0.0216 & -0.0022
\end{array}\right] }u_t \\
y_t &={\tiny\left[\begin{array}{llll}
0 & 1 & 0 & 0 \\
0 & 0 & 0 & 1
\end{array}\right]x_t}
\end{aligned}
\right.
\]
Three SS-to-ARX transformations are designed through different poles placement: the first one is based on the Cayley-Hamilton theorem from the subsection \ref{sec:CH-based SS-to-ARX transformation}, named as \textit{ARX-CH}, its ARX coefficients matrices are
\begin{equation*}
\begin{aligned}
&\overline{A}_1 = \footnotesize\left[\begin{array}{cc}
3.9944 & 0\\
0 & 3.9944
\end{array}\right], 
\overline{A}_2 = \footnotesize\left[\begin{array}{cc}
-5.8834 & 0\\
0 & -5.8834
\end{array}\right]\\
&\overline{A}_3 = \footnotesize\left[\begin{array}{cc}
3.7837 & 0\\
0 & 3.7837
\end{array}\right], 
\overline{A}_4 = \footnotesize\left[\begin{array}{cc}
-0.8947 & 0\\
0 & -0.8947
\end{array}\right]\\
&\overline{B}_1 = \footnotesize\left[\begin{array}{cc}
-0.0291 & -0.0143\\
-0.0216 & -0.0022
\end{array}\right],
\overline{B}_2 = \footnotesize\left[\begin{array}{cc}
0.0461 & 0.0386\\
0.0199 & 0.0012
\end{array}\right]\\
&\overline{B}_3 = \footnotesize\left[\begin{array}{cc}
-0.0049 & -0.0343\\
0.0213 & 0.0026
\end{array}\right],
\overline{B}_4 = \footnotesize\left[\begin{array}{cc}
-0.0121 & 0.0100\\
-0.0196 & -0.0016
\end{array}\right]
\end{aligned}
\end{equation*} 
the second and third ARX transformations are based on observer-theory from the subsection \ref{sec:observer-theory-based transformation}, and their poles are placed at $[0.01,0.02,0.03,0.04]$ and  $[0.04,0.08,0.12,0.16]$, named \textit{ARX-OT-1} and \textit{ARX-OT-2}, respectively. 

Based on their truncation errors, the order of  \textit{ARX-OT-1} and \textit{ARX-OT-2} are $4$ and $6$, respectively, and their ARX coefficient matrices are listed in Appendix \ref{appen_1}. 
We simulate three scenarios with different process and measurement noise magnitude, that is, $q=r=0, q=r=0.01,q=r=0.05$. The SS-based MPC cooperated with the Kalman filter with the given known noise covariances matrices, named \textit{SS-KF}. The input constraints are $|u_i| \leq 25^{\circ},i = 1, 2$, the output constraints are $-0.5\leq y_1 \leq 0.5 $ and $-100 \leq y_2 \leq 100$. The control goal is to make the pitch angle $y_2$ track a reference signal $r_2$. In designing the MPC controller we
take $W_y = \diag$([10,10]), $W_u = 0$, $W_{\Delta u}= \diag$([0.1, 0.1]), and the
prediction horizon is $T=10$.

\begin{figure*}[htbp]
\centering 
\subfigure[$q=r=0.0$, average MPC tracking cost $=8511.2422$]{
\begin{minipage}{8cm}\label{fig:1:a}
\centering
\includegraphics[scale=0.45]{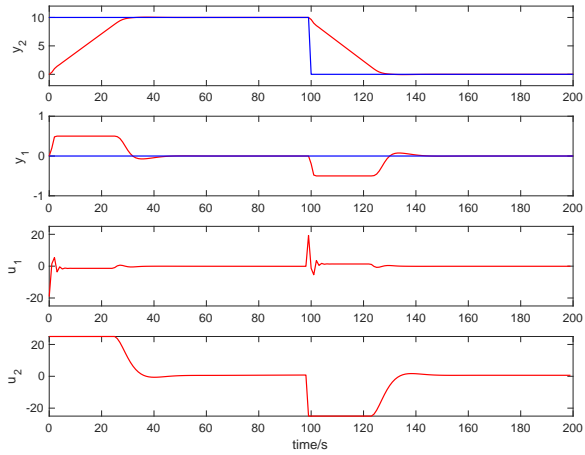}
\end{minipage}
}\subfigure[$q=r=0.01$, average MPC tracking cost $=26002.4804$]{
\begin{minipage}{8cm}\label{fig:1:b}
\centering
\includegraphics[scale=0.45]{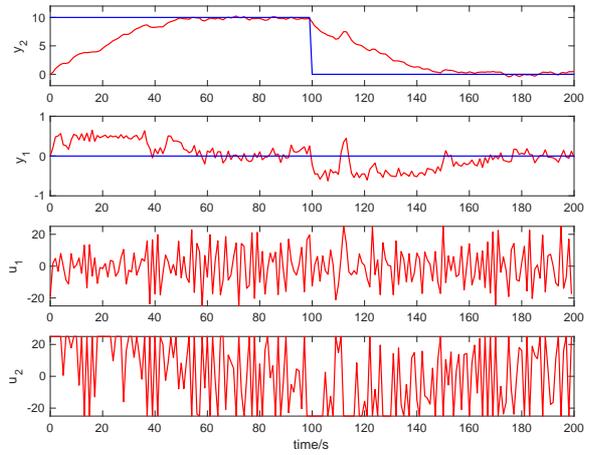}
\end{minipage}
}
\caption{The closed-loop control performance of \textit{ARX-CH}}
\label{fig:1}
\end{figure*}

\begin{figure*}[htbp]
\centering 
\subfigure[$q=r=0.01$, average MPC tracking cost $=8628.8547$]{   
\begin{minipage}{8cm}\label{fig:2:a}
\centering    
\includegraphics[scale=0.45]{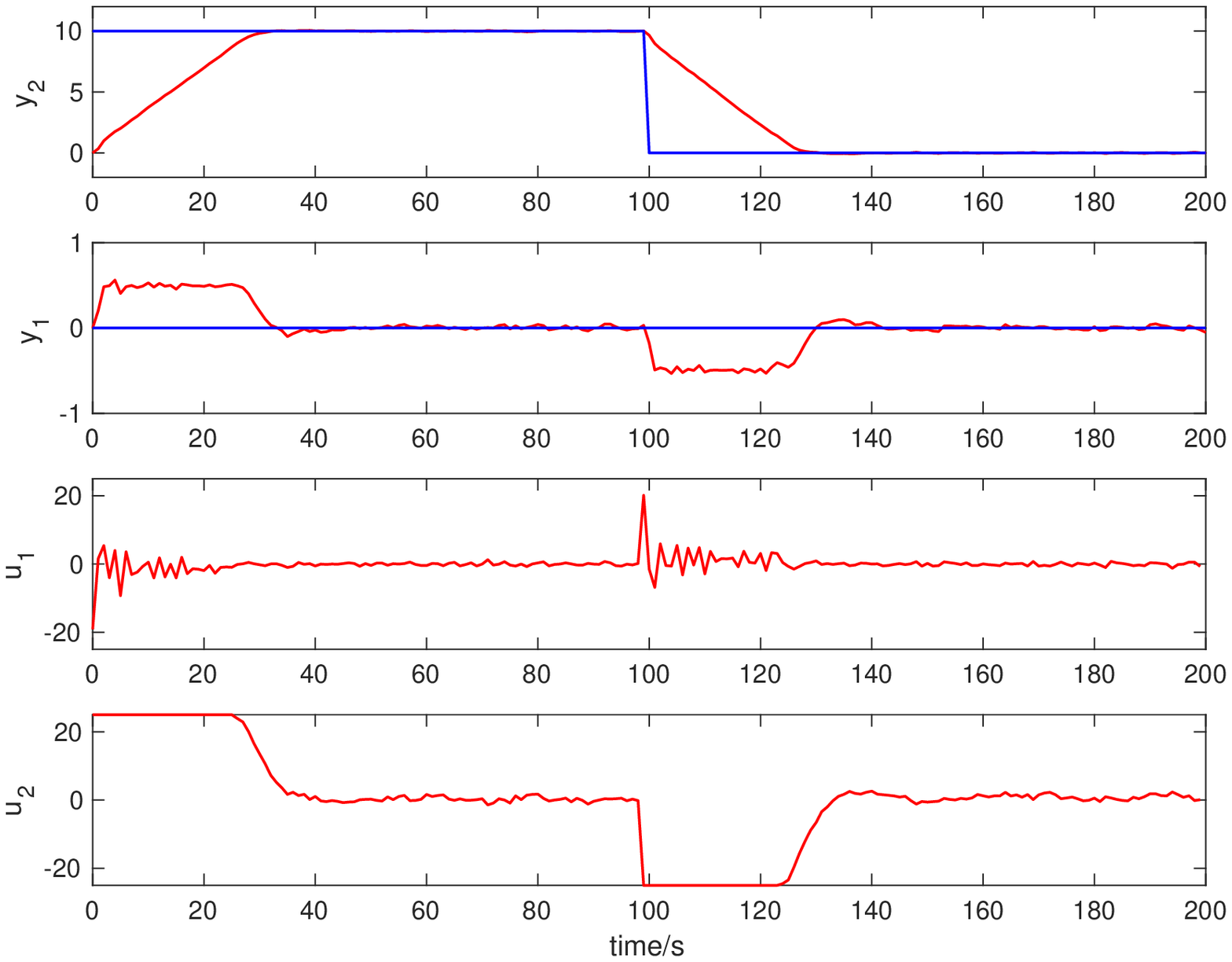}
\end{minipage}
}\subfigure[$q=r=0.05$, average MPC tracking cost $=9772.1471$]{
\begin{minipage}{8cm}\label{fig:2:b}
\centering
\includegraphics[scale=0.45]{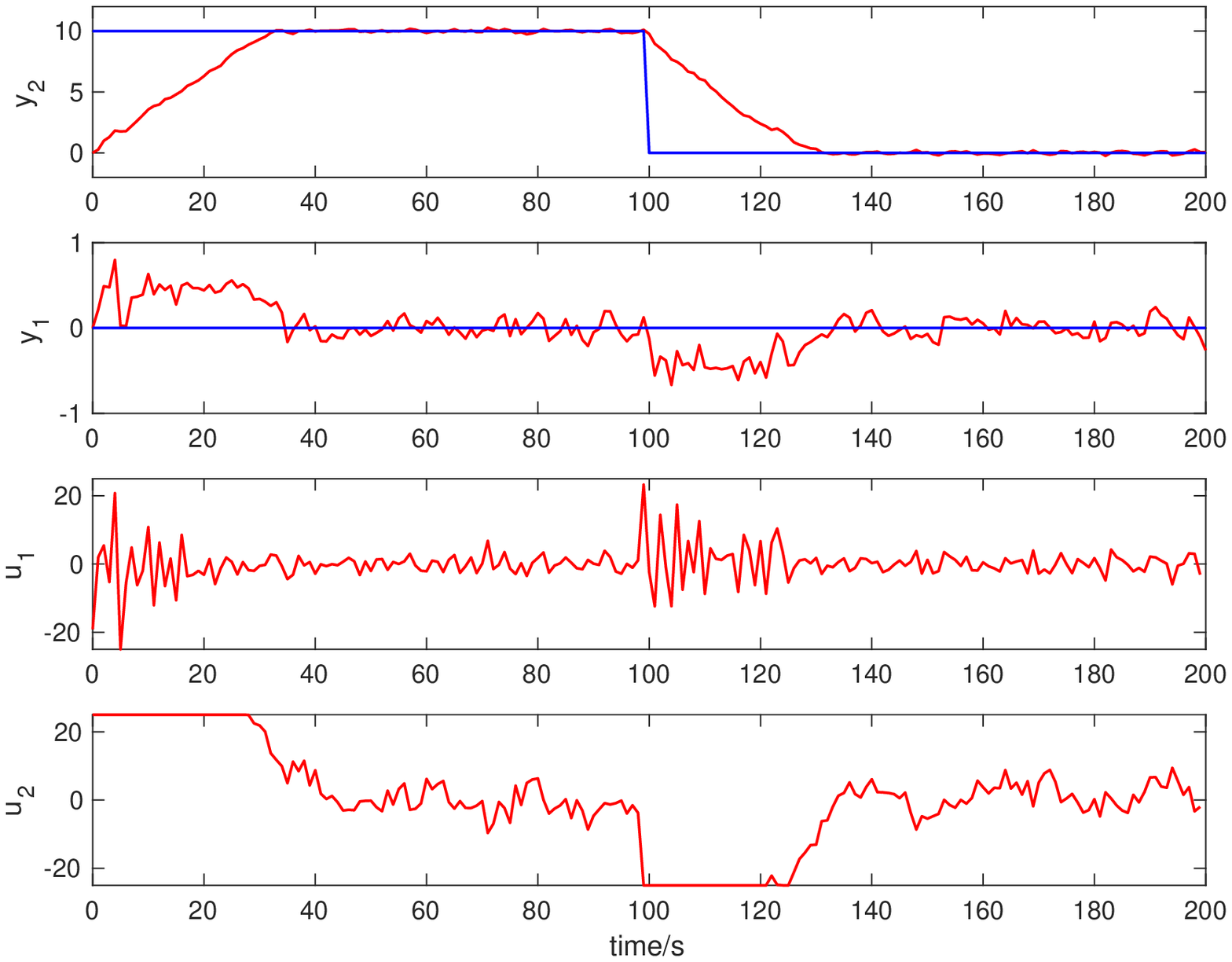}
\end{minipage}
}
\caption{The closed-loop control performance of \textit{ARX-Ob1}}
\label{fig:2}
\end{figure*}

\begin{figure*}[htbp]
\centering 
\subfigure[$q=r=0.01$, average MPC tracking cost $=8604.6714$]{
\begin{minipage}{8cm}\label{fig:3:a}
\centering
\includegraphics[scale=0.45]{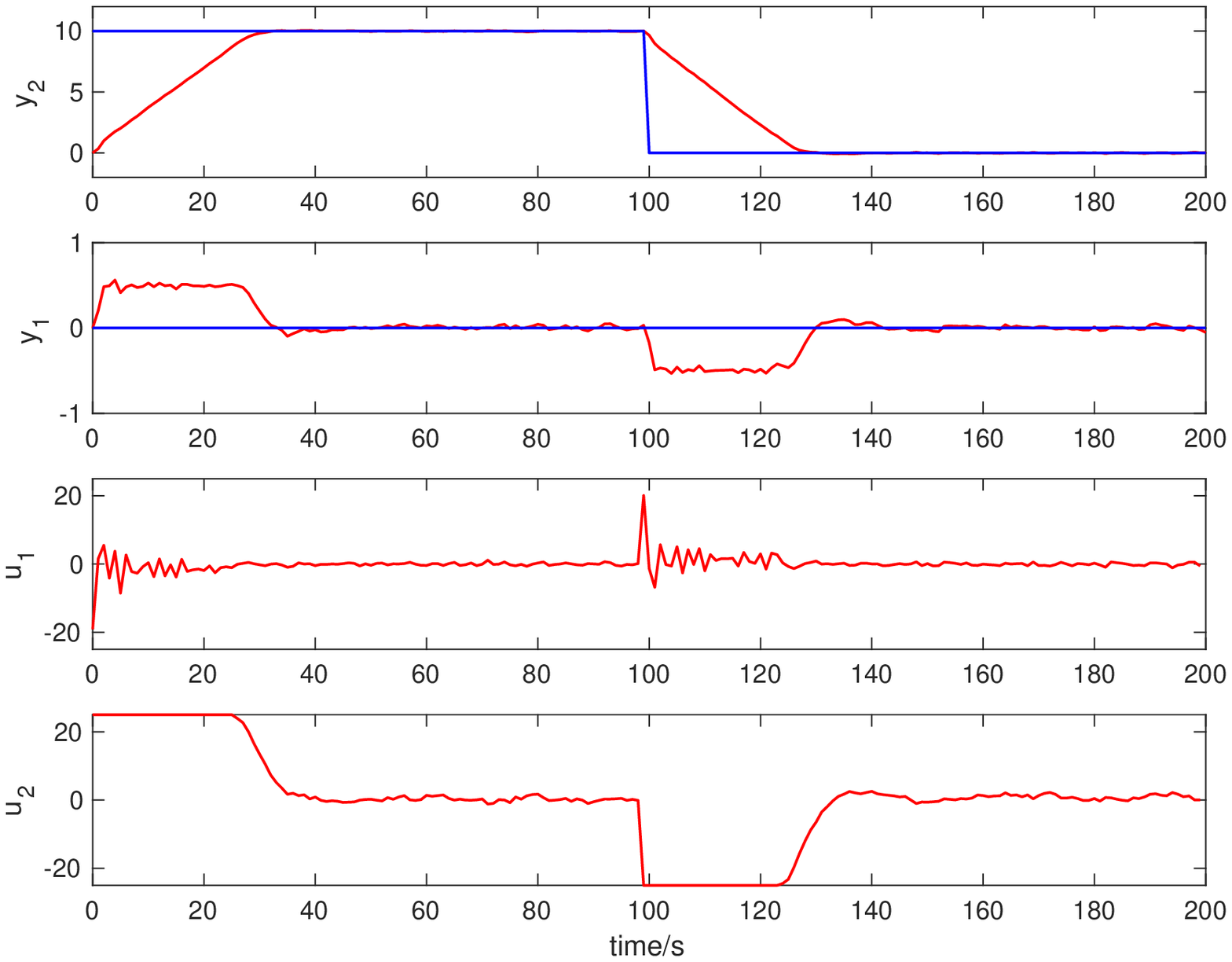}
\end{minipage}
}\subfigure[$q=r=0.05$, average MPC tracking cost $=9583.1624$]{
\begin{minipage}{8cm}\label{fig:3:b}
\centering
\includegraphics[scale=0.45]{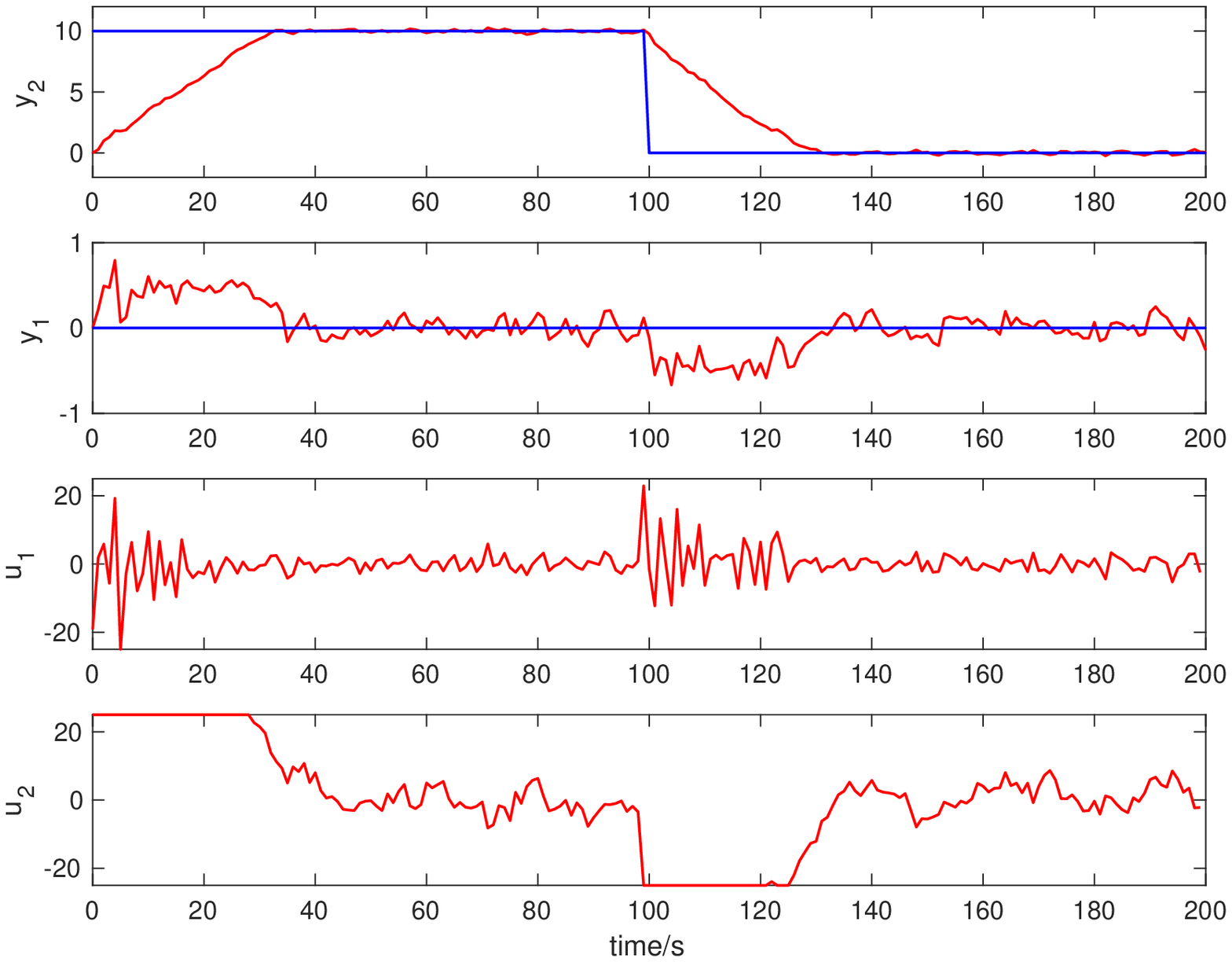}
\end{minipage}
} 
\caption{The closed-loop control performance of \textit{ARX-Ob2}}
\label{fig:3}
\end{figure*}

\begin{figure*}[htbp]
\centering 
\subfigure[$q=r=0.01$, average MPC tracking cost $=8596.0559$]{
\begin{minipage}{8cm}\label{fig:4:a}
\centering
\includegraphics[scale=0.45]{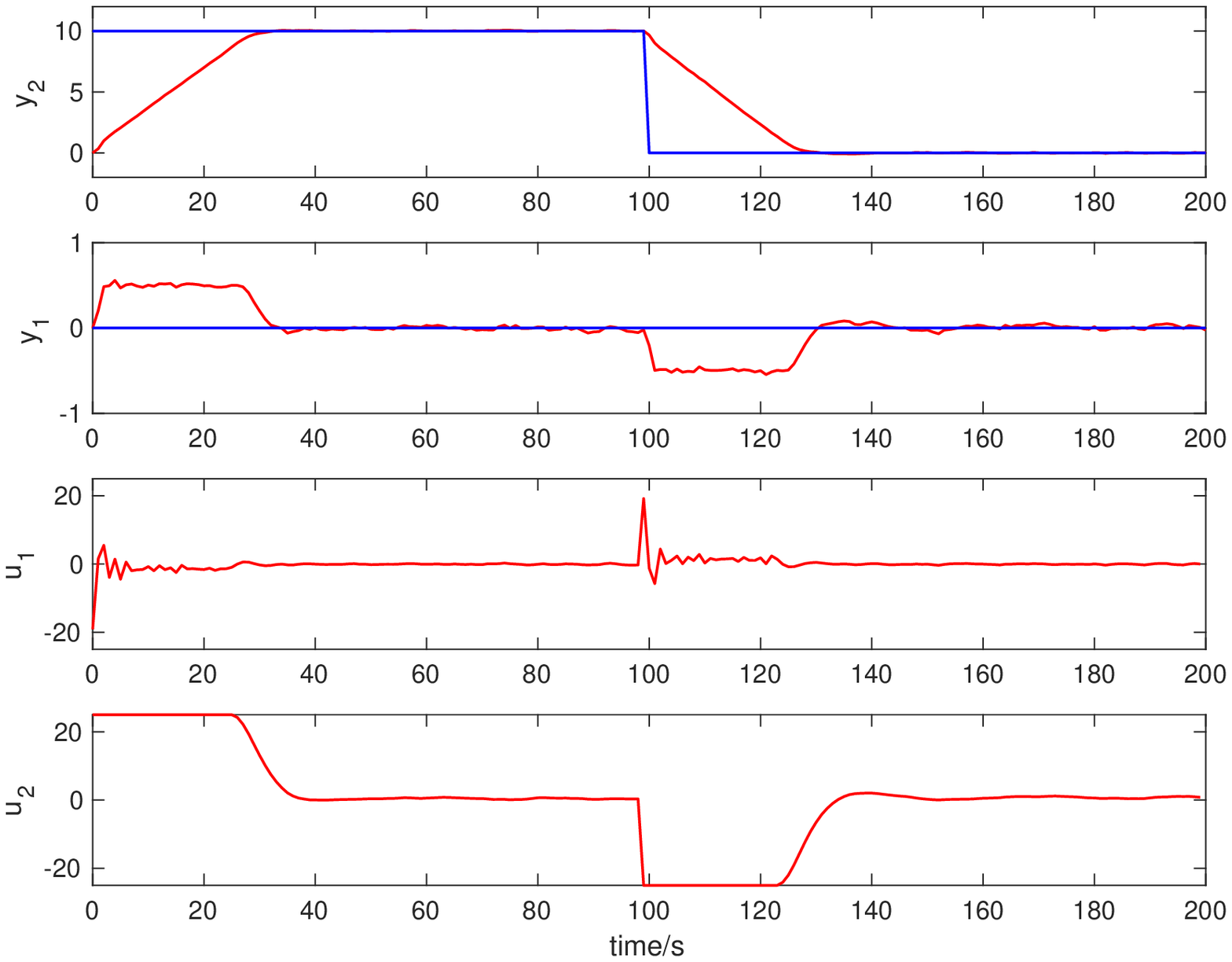}
\end{minipage}
}\subfigure[$q=r=0.05$, average MPC tracking cost $=9043.3026$]{
\begin{minipage}{8cm}\label{fig:4:b}
\centering 
\includegraphics[scale=0.45]{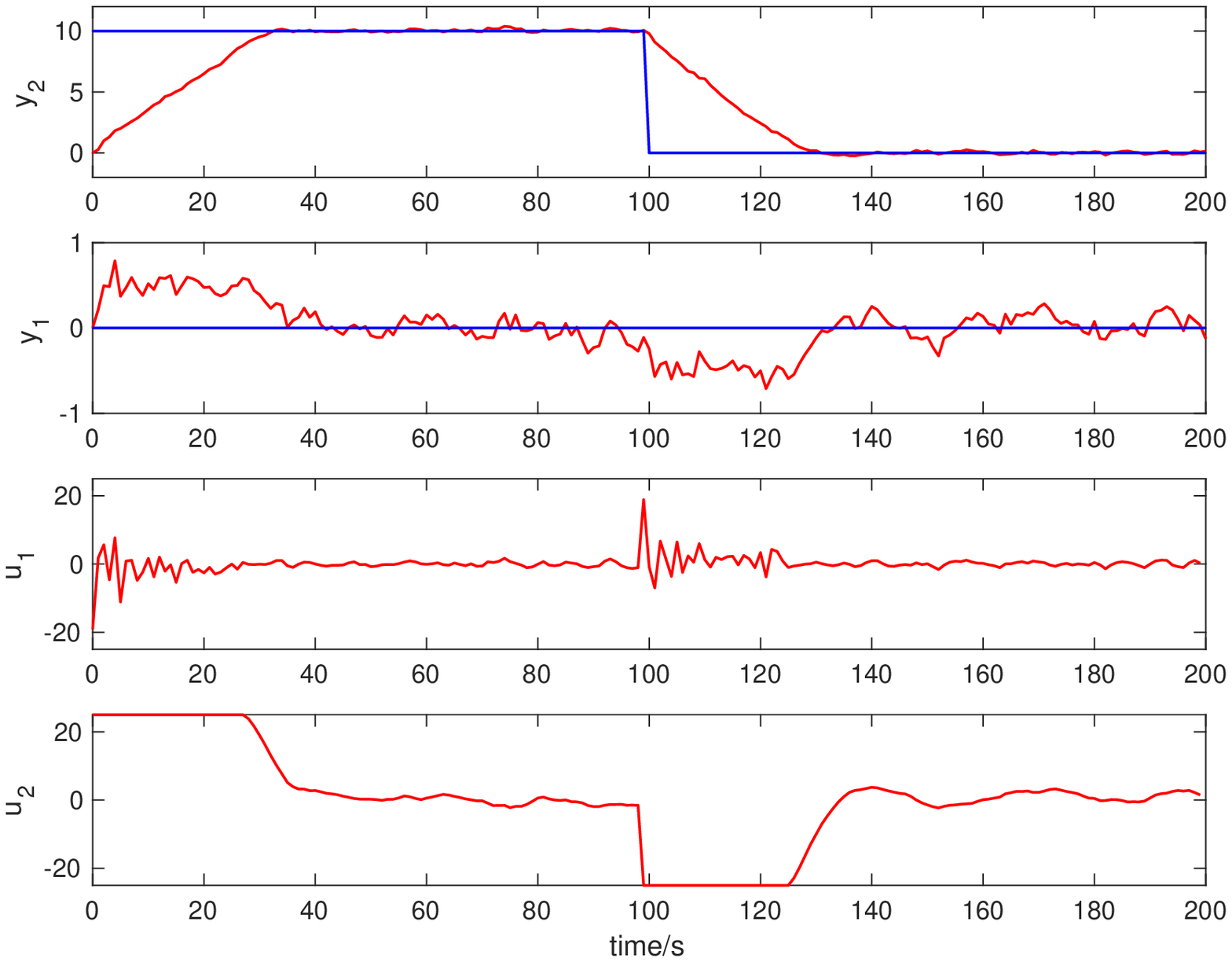}
\end{minipage}
}
\caption{The closed-loop control performance of \textit{SS-KF}}
\label{fig:4}
\end{figure*}

Fig. \ref{fig:1:a} is the closed-loop performance results of \textit{ARX-CH} under $q=r=0$, namely without any noise. The \textit{ARX-OT-1} and \textit{ARX-OT-2} and \textit{SS-KF} transformations all coincide with the same plot results (thus omitted), which shows that the tracking performance is good and there are no constraint violations in input and output. It tells the equivalence of SS-based and ARX-based MPC problems under the presented SS-to-ARX transformations. But the \textit{ARX-CH} transformation suffers bad closed-loop performance under noise $q=r=0.01$ see Fig. \ref{fig:1:b}, and diverges under noise $q=r=0.05$ (thus no figure). It shows that the CH-based transformation is very sensitive to noise. The Fig. \ref{fig:2}, \ref{fig:3} and \ref{fig:4} are the closed-loop control simulation results of \textit{ARX-OT-1}, \textit{ARX-OT-2} and \textit{SS-KF}, under noise $q=r=0.01$ and $q=r=0.05$, respectively. Clearly, they all are more robust than \textit{ARX-CH}, and the order of better control performance is \textit{ARX-OT-1} and \textit{ARX-OT-2} and \textit{SS-KF}, which is represented by the average MPC tracking cost. It also shows that the larger values of placed poles can be more robust to noise, and the closed-loop performance can approximate the optimal Kalman Filter but at the cost of the required larger order of the ARX model.

\section{Conclusion}\label{sec:conclusion}
This paper proposes the first-principle-based modeling ARX acquiring paradigm. This paradigm first obtains control-oriented SS models through linearizing the first-principle-based models, which are then transformed into equivalent ARX models via the Cayley-Hamilton, Observer-Theory, and Kalman Filter based SS-to-ARX transformations presented in this paper. Numerical results of the AFTI-16 MPC numerical example show the equivalence of SS-based and ARX-based MPC problems and the noise robustness of different SS-to-ARX transformations, which point out that choosing the ARX model order should depend on the process and measurement noise, to achieve a good closed-loop performance, rather than depending on fitting criteria in data-driven ARX identification paradigm. 

This interpretative ARX model can naturally be adopted in an adaptive MPC framework by adding an online updating scheme for the ARX model, such as our work in \cite{wu2022interpretative}. In addition, the SS-to-ARX transformation can also be used as the basis for another future direction, which is transforming a nonlinear state-space model into an equivalent nonlinear ARX model. By sampling the space of the input and state, a set of linear parameter-varying (LPV) SS models can be generated and transformed into their equivalent LPV-ARX models, which are then merged into a nonlinear ARX model, to achieve model reduction.

\bibliographystyle{unsrt}
\bibliography{ref} 

\begin{appendices}
\section{SS-to-ARX transformation}\label{appen_1}
The ARX coefficient of \textit{ARX-OT-1} transformation  are listed as follows,
\begin{equation*}
\begin{aligned}
&\overline{A}_1 = \footnotesize\left[\begin{array}{cc}
1.8998 & 0.0525\\
0.0265 & 1.9946
\end{array}\right], 
\overline{A}_2 = \footnotesize\left[\begin{array}{cc}
-0.7583 & -0.0584\\
0.0647 & -0.9442
\end{array}\right]\\
&\overline{A}_3 = \footnotesize\left[\begin{array}{cc}
-0.0389 & 0.0053\\
0.0128 & -0.0483
\end{array}\right],\\
&\overline{A}_4 = \footnotesize\left[\begin{array}{cc}
-0.0016 & 5.0465\times10^{-4}\\
8.4386\times10^{-4} & -0.0020
\end{array}\right],\\
&\overline{B}_1 = \footnotesize\left[\begin{array}{cc}
-0.0291 & -0.0143\\
-0.0216 & -0.0022
\end{array}\right],
\overline{B}_2 = \footnotesize\left[\begin{array}{cc}
-0.0138 & 0.0088\\
-0.0225 & -0.0028
\end{array}\right],\\
&\overline{B}_3 = \footnotesize\left[\begin{array}{cc}
-4.8002\times10^{-4} & -4.6250\times10{-4}\\
-9.7441\times10^{-4} & -2.4731\times10^{-4}
\end{array}\right],\\
&\overline{B}_4 = \footnotesize\left[\begin{array}{cc}
-1.2950\times10^{-5} & 1.9723\times10^{-5}\\
-3.4876\times10^{-5} & -1.3766\times10^{-5}
\end{array}\right].
\end{aligned}
\end{equation*} 

The ARX coefficient of \textit{ARX-OT-2} transformation are listed as follows,
\begin{equation*}
\begin{aligned}
&\overline{A}_1 = \footnotesize\left[\begin{array}{cc}
1.7461 & 0.0788\\
0.0606 & 1.8483
\end{array}\right], 
\overline{A}_2 = \footnotesize\left[\begin{array}{cc}
-0.4909 & -0.0955\\
0.0132 & -0.6762
\end{array}\right],\\
&\overline{A}_3 = \footnotesize\left[\begin{array}{cc}
-0.1156 & 0.0096\\
0.0328 & -0.1443
\end{array}\right],\\
&\overline{A}_4 = \footnotesize\left[\begin{array}{cc}
-0.0201 & 0.0055\\
0.0094 & -0.0235
\end{array}\right],\\
&\overline{A}_5 = \footnotesize\left[\begin{array}{cc}
-0.0033 & 0.0013\\
0.0020 & -0.0036
\end{array}\right], \\
&\overline{A}_6 = \footnotesize\left[\begin{array}{cc}
-5.1508\times10{-4} & 2.6754\times10{-4}\\
3.5971\times10{-4} & -5.4367\times10{-4}
\end{array}\right],\\
&\overline{B}_1 = \footnotesize\left[\begin{array}{cc}
-0.0291 & -0.0143\\
-0.0216 & -0.0022
\end{array}\right],
\overline{B}_2 = \footnotesize\left[\begin{array}{cc}
-0.0177 & 0.0066\\
-0.0247 & -0.0026
\end{array}\right],\\
&\overline{B}_3 = \footnotesize\left[\begin{array}{cc}
-0.0025 & 0.0016\\
-0.0040 & -8.6513\times10^{-4}
\end{array}\right],\\
&\overline{B}_4 = \footnotesize\left[\begin{array}{cc}
-3.0362\times10^{-4} & 2.7747\times10^{-4}\\
-5.4133\times10^{-4} & -1.8877\times10^{-4}
\end{array}\right],\\
&\overline{B}_5 = \footnotesize\left[\begin{array}{cc}
-3.3682\times10^{-5} & 4.5849\times10^{-5}\\
-7.0212\times10^{-5} & -3.5253\times10^{-5}
\end{array}\right],\\
&\overline{B}_6 = \footnotesize\left[\begin{array}{cc}
-3.5278\times10^{-6} & 7.3794\times10^{-6}\\
-9.0504\times10^{-6} & -6.1389\times10^{-6}
\end{array}\right].
\end{aligned}
\end{equation*} 
\end{appendices}
\end{document}